\documentclass[12pt, JNE]{iopart}

\newcolumntype{H}{>{\setbox0=\hbox\bgroup}c<{\egroup}@{}}
\DeclareMathOperator*{\argmax}{arg\,max}

\begin{document}

\pagenumbering{gobble}
This is the version of the article before peer review or editing, as submitted by an author to Journal of Neural Engineering. IOP Publishing Ltd is not responsible for any errors or omissions in this version of the manuscript or any version derived from it. The Version of Record is available online at \url{https://doi.org/10.1088/1741-2552/aae8c7}.
\newpage\pagenumbering{arabic}
	
\title{Direct information transfer rate optimisation for SSVEP-based BCI}

\author{Anti Ingel}
\address{Institute of Computer Science, University of Tartu, Tartu, Estonia}
\ead{antiingel@gmail.com}

\author{Ilya Kuzovkin}
\address{Institute of Computer Science, University of Tartu, Tartu, Estonia}
\ead{ilya.kuzovkin@gmail.com}

\author{Raul Vicente}
\address{Institute of Computer Science, University of Tartu, Tartu, Estonia}
\ead{raulvicente@gmail.com}
\vspace{10pt}
\begin{indented}
\item[]June 2018
\end{indented}

\begin{abstract}
In this work, a classification method for SSVEP-based BCI is proposed. The classification method uses features extracted by traditional SSVEP-based BCI methods and finds optimal discrimination thresholds for each feature to classify the targets. Optimising the thresholds is formalised
as a maximisation task of a performance measure of BCIs called information transfer rate (ITR). However, instead of the standard method of calculating ITR, which makes certain assumptions about the data, a more general formula is derived to avoid incorrect ITR calculation when the standard assumptions are not met. This allows the optimal discrimination thresholds to be automatically calculated and thus eliminates the need for manual parameter selection or performing computationally expensive grid searches.

The proposed method shows good performance in classifying targets of a BCI, outperforming previously reported results on the same dataset by a factor of 2 in terms of ITR. The highest achieved ITR on the used dataset was 62 bit/min. The proposed method also provides a way to reduce false classifications, which is important in real-world applications.
\end{abstract}

%
%
%
%
%

\section{Introduction}

In this work, a classification method for steady-state visual evoked potential (SSVEP) based brain-computer interface (BCI) is proposed. Brain-computer interface is a direct communication channel between the brain and some external device. SSVEP-based BCI uses visual stimuli, called targets, to elicit a response in the brain of the user and then extract features from the recorded brain activity and, finally, classifies the result as one of the commands that the user can send to the external device.

Often, the visual stimuli used in SSVEP-based BCIs are multiple squares displayed on a computer screen, each flickering with different frequency~\cite{ssvep_stim}. Different flickering frequencies elicit different brain responses and there are many feature extraction methods to evaluate the presence of this signal in a recorded electroencephalography (EEG) signal.

The novelty of the classification method proposed in this work is that it is based on information transfer rate (ITR) maximisation. ITR is a standard measure of performance for BCIs~\cite{itr_wolpaw}. It combines the accuracy and the speed of the classifier into a single number which shows how much information is transferred by the BCI in one unit of time. Therefore, maximising ITR maximises the amount of information that the BCI can transfer in fixed time interval.

Often, for SSVEP-based BCIs with similar classification rules, parameters like discrimination threshold are either chosen manually or found using grid search approaches~\cite{11_fft_bilinear,12_amuse_ica,1_music,4_fft_svm,7_bifb_psda_cca}. This work proposes classification method that finds discrimination thresholds so that they maximise ITR thus improving the performance of the BCI. It is faster than grid-search approach and in addition provides a way to compare different feature extraction methods and parameters of the BCI without having to worry that the results are affected by the manual choice of discrimination threshold.

The proposed method also provides a way to combine multiple feature extraction methods using linear discriminant analysis (LDA). LDA is used to project feature vectors produced by multiple feature extraction methods into lower dimensional space after which the classification algorithm can be used the same way as for one feature extraction method with distances to the decision borders as new features. Combining multiple feature extraction methods together with the threshold selection described above, the present approach achieves more than two times larger ITR than previously reported comparable result. The comparison was made to the works in which the same SSVEP dataset was used as in this work to make the comparison as fair as possible.

\section{Methods}
\subsection{SSVEP dataset for offline experiments}

The performance of the proposed method was tested using publicly available SSVEP dataset by Bakardijan \textit{et al.}\footnote{\url{http://www.bakardjian.com/work/ssvep_data_Bakardjian.html}} which contains SSVEP data from four subjects~\cite{dataset}. The dataset contains EEG recordings of the subjects' brain responses to 8 Hz, 14 Hz and 28 Hz flickering. The visual stimuli were displayed on a computer monitor with 170 Hz refresh rate, placed approximately 90 cm away from the subject’s eyes.

For each subject, the dataset contains five trials for each flickering frequency. Each trial consists of about 25 seconds of EEG data, 10 seconds of which is without the visual stimulation. EEG data in the dataset was recorded using BIOSEMI EEG system with 128 channels and 256 Hz sampling rate. However, in this work only the data from two channels---O1 and O2---were used. These channels were chosen to make the results more comparable to the previous works in which similar channels were used (see Section~\ref{comparison}). The exact choice of O1 and O2 was made to increase comparability to the results where consumer-grade EEG device Emotiv EPOC was used and for Emotiv EPOC, channels O1 and O2 are most suitable for SSVEP detection.

\subsection{Feature extraction methods}
In this work, two feature extraction methods were used
\begin{itemize}
	\item Power spectral density analysis (PSDA)~\cite{psda}
	\item Canonical correlation analysis (CCA)~\cite{cca}
\end{itemize}
Both of these methods extract a feature for each target, that is for every possible command of the BCI. In general, the larger the feature value for a target, the more likely the target is the user's choice. The same holds for other feature extraction methods, like Minimum energy combination~\cite{mec}, Likelihood ratio test~\cite{LRT} and Continuous wavelet transform~\cite{cwt}. Thus these feature extraction methods could also be used with the proposed classification method. 

\subsubsection{Power spectral density analysis}

Power spectral density analysis (PSDA)~\cite{psda} feature extraction method is based on estimating the power spectrum of the EEG signal using fast fourier transform (FFT). Since each of the flickering targets elicits brain response with the same fundamental frequency as the flickering itself, the estimated power of this frequency serves as a good feature for classifying which of the targets is the user looking at. But since brain signals are inherently very noisy, single power value at certain time point is not enough to reliably classify the target. Thus the feature is calculated from a fixed-size signal segment after every fixed time step. In the experiments of this paper, window length of 1 second and time step of 0.125 seconds were used. In addition to the power of fundamental frequency of the flickering, also the powers of its second and third harmonic were used as features. To obtain one feature per target, all the powers obtained for one target can be added together. However, in the experiments of this work, LDA dimensionality reduction, described in Section \ref{lda_section}, was used to find one feature for each target. These features are then used to classify the target. 

\subsubsection{Canonical correlation analysis}

Canonical correlation analysis (CCA)~\cite{cca} feature extraction method is inherently multidimensional, meaning that it calculates single feature value for each target for multichannel EEG signal. This is not the case with the PSDA method, which calculates powers for each channel separately. CCA uses canonical correlation as a feature value, which is a correlation between two sets of signals---one set is the multichannel EEG signal and the other is set of reference signals, which mimic the expected behaviour of the brain's response to the stimulation. Each target has its own set of reference signals and reference signals also contain signals for the second and third harmonic of the flickering frequency. The feature extraction method calculates canonical correlation of each set of reference signals to the multichannel EEG signal. As before, the larger the correlation is, the more likely the corresponding target is the user's choice.

\subsection{Combining multiple feature extraction methods}\label{lda_section}

To make it possible for the BCI to use multiple feature extraction methods at the same time and thus improve performance of the BCI, a dimensionality reduction and classification method called linear discriminant analysis (LDA) was used. LDA projects the input data to a lower-dimensional space so that the variance between the classes is maximised with respect to the variance within the classes. To classify a new sample, LDA projects the new sample to the lower-dimensional space and finds which class mean is closest to it. This classification rule defines decision borders for the classes and the distance from the decision border to a new sample can be interpreted as confidence score for corresponding class. Therefore, to combine multiple feature extraction methods, LDA is used to find the lower-dimensional space into which the original features are projected. Then their distances to decision borders are used as new features. This combines the information from all the original features into single new feature per target. Having one feature per target is needed because this is required by the classification rule described in section~\ref{class_rule}.

\section{Results}

\subsection{Classification method}

This section describes the derivation of the proposed classification method that is based on ITR maximisation. It follows the description of the method given in~\cite{magister}. First, the classification rule is described, and then a formula for estimating ITR from discrimination thresholds is derived. Finally, it is shown that the formula is differentiable and thus gradient descent can be used to find optimal discrimination thresholds. For more details about the derivation see \ref{appendix_conditioning} and \ref{appendix_gradient}. Implementation of the algorithm can be found in \ref{appendix_repo}.

\subsubsection{Classification rule}\label{class_rule}

Assuming that there are $n$ targets $\{1,\dots,n\}$, a feature extraction method extracts $n$ features $\{f_1,\dots,f_n\}$. The classification rule uses a discrimination threshold $t_j$ for each feature $f_j$ to classify a sample as follows:
\begin{tcolorbox}
	A sample is classified as class $k$ if for this sample $f_k\geq t_k$ and $f_j < t_j$ for $j\in\{1,2\dots,n\}\setminus\{k\}$.
\end{tcolorbox}
\noindent Or in other words, a sample is classified as class $k$ if the feature value for this class is over the threshold and all the other feature values are below their thresholds.

It can be seen that this rule might not classify all the samples. If multiple features are over their thresholds, or none of the features are, then classification is not made. This case is viewed as the BCI not being confident enough and thus it avoids making a prediction. This is useful when not making a prediction is less costly than making a wrong prediction.

\subsubsection{Finding the thresholds}\label{thresh}

Finding the thresholds $\{t_1,\dots,t_n\}$ will be formalised as a maximisation task, where the function to be maximised is a performance measure of the BCI. The performance measure $\mbox{ITR}(\vec{t}\,)$ is a function of thresholds $\vec{t}=(t_1,\dots,t_n)^T$ and thus the maximisation task is
\begin{align}
\argmax_{\vec{t}}\mbox{ITR}(\vec{t}\,)
\end{align}
which was solved in this work using gradient descent algorithm. The performance measure itself is defined similarly to standard ITR, which is used to evaluate SSVEP-based BCIs, but the assumptions~\cite{itr_wolpaw} about precisions, the probabilities of errors and predictions is not made. In particular, the standard assumptions are
\begin{itemize}
	\item all predictions are equally likely
	\item if choosing a wrong target, all wrong targets are equally likely to be
	chosen
	\item the accuracy of successfully classifying a target has to be the same for every
	target.
\end{itemize}
\emph{Making these assumptions while they are not met can result in incorrect ITR calculation} and can cause the maximisation algorithm to return non-optimal thresholds. 

The performance measure is desired to be a continuous function of thresholds and will be defined as
\begin{align}\label{itr_min}
\mbox{ITR}(\vec{t}\,)=\mbox{MI}(\vec{t}\,)\cdot\frac{60}{\mbox{MDT}(\vec{t}\,)}
\end{align}
where $\mbox{MI}(\vec{t}\,)$ is the mutual information between random variables $P$ and $C$, which model the predicted class and the correct class respectively, and $\mbox{MDT}(\vec{t}\,)$ is the mean detection time in seconds. Mutual information $\mbox{MI}(\vec{t}\,)$ shows how many bits of information is transferred with one prediction and is multiplied with the amount of predictions made in one minute, thus giving units of bits per minute.

\subsubsection{Calculating mutual information}

By denoting the event of predicting class $i$ by $P_i$ and the event of correct class being $j$ by $C_j$, the mutual information between random variables $P$ and $C$ as defined previously is
\begin{align}\label{itr}
\mbox{MI}(\vec{t}\,)=I(P, C)=\sum_{i=1}^{n}\sum_{j=1}^{n}\mathbf{P}(P_i\cap C_j)\log_2\left(\frac{\mathbf{P}(P_i\cap C_j)}{\mathbf{P}(P_i)\cdot\mathbf{P}(C_j)}\right).
\end{align}
When using the assumptions discussed in \ref{thresh} the formula \eqref{itr} can be simplified to standard ITR~\cite{itr_wolpaw} calculation formula in units of bits per prediction
\begin{align}\label{itr_standard}
\mbox{ITR}_{S}=\log_2N+P\log_2P(1-P)\log_2\left(\frac{1-P}{N-1}\right)
\end{align}
where $N$ is the number of targets and $P$ is the classification accuracy. Therefore the proposed ITR calculation method is a generalisation of the standard method.

Next we show that the mutual information is indeed a function of thresholds. If the feature value for class $k$ is modelled with random variable $F_k$, the event of predicting class $i$ can be represented as
\begin{align}
P_i=\left\{\omega\in\Omega: F_i(\omega)\geq t_i\land\left(\bigwedge\limits_{j\in\{1,\dots,n\}\setminus\{i\}}F_j(\omega)<t_j\right)\right\}
\end{align}
where $\Omega$ denotes the sample space. From this, by making the assumption that the features are conditionally independent, conditioned on the correct class $C$, the conditional probability of predicting a class can be calculated as
\begin{align}
\mathbf{P}(P_i\mid C_k)=\left(1-{F}_{F_i\mid C_k}(t_i)\right)\prod_{j\in\{1,\dots,n\}\setminus\{i\}}F_{F_j\mid C_k}(t_j)
\end{align}
where ${F}_{F_\ell\mid C_k}$ denotes the cumulative distribution function (CDF) of the conditional distribution $F_\ell\mid C_k$. It was empirically observed in this work that skew normal distribution models well the conditional distributions of features (see \ref{distribution}). Thus by using least squares method for fitting skew normal distribution to the data, the required CDFs can be calculated.

Therefore, after having fit skew normal distributions to the data, the probability $\mathbf{P}(P_i\mid C_k)$ only depends on the thresholds $\vec{t}$. Now, by using known rules from probability theory, all the probabilities required in calculating mutual information can be estimated
\begin{align}
\mathbf{P}(C_j)&=\frac{\mbox{Number of samples from class $j$}}{\mbox{Number of samples}}\label{Cj}\\
\mathbf{P}(P_i)&=\sum_{j=1}^{n}\mathbf{P}(P_i\mid C_j)\mathbf{P}(C_j)\label{Pi}\\
\mathbf{P}(P_i\cap C_j)&=\mathbf{P}(P_i\mid C_j)\mathbf{P}(C_j).
\end{align}
Thus mutual information can indeed be represented as a continuous function of thresholds and hence its gradient can be calculated.

\subsubsection{Calculating mean detection time}

So far it has been shown that the amount of information transferred with one prediction can be calculated as a function of thresholds. Now we show that the mean detection time can also be represented as function of thresholds. 

Since it is desired that the performance measure would reflect the performance of the online BCI, a method for estimating online MDT is proposed. First, let us discuss how the implemented BCI splits the incoming signal. The feature extraction methods always use $w$ seconds of data at a time. Whether there is an overlap between consecutive windows depends on whether the last prediction was successful, that is prediction was made. If a prediction is made, there is no overlap between windows. In this case the time step between consecutive feature extractions is $w$. If a prediction is not made, there is $w-s$ seconds of overlap between windows. In this case $s$ is the time step between consecutive feature extractions.

By making the assumption that the probability of making a prediction is the same in every time step and that the probability of making a prediction will be the same in online case as it was offline, the MDT can be calculated as
\begin{align}
\mbox{MDT}(\vec{t}\,)=w+\left(\frac{1}{\mathbf{P}\left(\bigcup_{i=1}^nP_i\right)}-1\right)\cdot s
\end{align}
where $s$ is multiplied with the expected number of failures before a successful prediction.

MDT is a continuous function of thresholds, because 
\begin{align}
\mathbf{P}\left(\bigcup_{i=1}^nP_i\right)=\sum_{i=1}^{n}\mathbf{P}(P_i)
\end{align}
and we have already shown that $\mathbf{P}(P_i)$ is a continuous function of thresholds. Thus the performance measure as shown in equation \eqref{itr_min} is a continuous function of thresholds.

\subsubsection{Gradient Descent}

Finally, thresholds which maximise the ITR directly are found using gradient ascent (GA) algorithm. GA iteratively updates the threshold values by moving in the direction of greatest increase of ITR using the rule
\begin{align}
\vec{t}_{n+1}=\vec{t}_n +\mu\cdot\nabla\mbox{ITR}(\vec{t}_n)
\end{align}
where $\nabla\mbox{ITR}$ is the gradient of ITR and $\mu$ is step size parameter. For the calculation of the gradient, see \ref{appendix_conditioning} and \ref{appendix_gradient}. GA stops if ITR improves less than $10^{-6}$. Since $\mbox{ITR}(\vec{t}\,)$ is not concave, GA has to be run multiple times, each time obtaining a local maximum and the final result will be the largest from these local maxima. In this work, 20 random initialisations were used for GA.

\subsection{Comparison of the proposed method to previous results}\label{comparison}

The performance of the proposed method was tested using SSVEP dataset by Bakardijan \textit{et al.}~\cite{dataset} and the results were compared to two other works~\cite{7_bifb_psda_cca,11_fft_bilinear} where the same dataset was used. Detailed comparison of these works can be seen in~\ref{related_work}.

In the work by Demir \textit{et al.}~\cite{7_bifb_psda_cca}, a feature extraction method called bio-inspired filter bank (BIFB) is introduced. The method is in essence very similar to PSDA method. In their method, power spectral density of the EEG signal is estimated and triangular filters are used on the result to make the powers of different frequencies more comparable. They also compare their method to CCA feature extraction method. Only the data for Oz electrode is used in their work.

In the work by Jukiewicz and Cysewska-Sobusiak~\cite{11_fft_bilinear} PSDA method was used for feature extraction and two classification methods were compared---bilinear separation and SVM. They only used electrodes O1, O2 and Oz and two classes 8 Hz and 14 Hz.

These two works were chosen, because similarly to this work, the classifier was allowed to not make predictions for some samples. As can be seen from Table~\ref{tab:classification}, the possibility to not make predictions results in better performance. Overview of the results of other works where the same dataset was used for testing SSVEP-based BCI can be seen in~\ref{related_work}. 

The proposed classifier was also compared to Random Forest to see how well it performs compared to standard machine learning algorithm. Detailed comparison can be seen in Table~\ref{tab:classification} and comparison to other works and to random forest can be seen in Table~\ref{table}. All the tables contain both, the standard ITR and proposed ITR. The difference between them is that standard ITR uses formula \eqref{itr_standard} to estimate the amount of information sent by one prediction, while proposed ITR uses formula \eqref{itr}. Table~\ref{tab:extraction} contains the comparison of PSDA and CCA feature extraction methods to their combination. In this work, classifiers were evaluated using five-fold cross-validation, each fold corresponding to data from separate recording trial. 

\begin{table}[]
	\def\arraystretch{1.5}
	\centering
	\caption{Comparison of CCA, PSDA feature extraction methods and their combination, using proposed classifier.}
	\label{tab:extraction}
	\resizebox{\textwidth}{!}
	{
	\begin{tabular}{|l|l|l|l|l!{\vrule width 2pt}l!{\vrule width 2pt}l|l|l|l!{\vrule width 2pt}l!{\vrule width 2pt}l|l|l|l!{\vrule width 2pt}l!{\vrule width 2pt}}
		\hline
		& \multicolumn{5}{l|}{\textbf{CCA + PSDA (using LDA)}}                 & \multicolumn{5}{l|}{\textbf{CCA}}                                    & \multicolumn{5}{l|}{\textbf{PSDA}}                                   \\ \cline{2-16} 
		\multirow{-2}{*}{\textbf{}} & \textbf{S1} & \textbf{S2} & \textbf{S3} & \textbf{S4} & \textbf{Avg} & \textbf{S1} & \textbf{S2} & \textbf{S3} & \textbf{S4} & \textbf{Avg} & \textbf{S1} & \textbf{S2} & \textbf{S3} & \textbf{S4} & \textbf{Avg} \\ \hline
		\rowcolor[HTML]{EFEFEF} 
		\textbf{ITR proposed}               & 64.62       & 53.73       & 33.19       & 5.65        & \textbf{39.30}        & 55.6        & 47.33       & 13.71       & 14.47       & \textbf{32.78}        & 43.41       & 28.2        & 16.68       & 5.06        & \textbf{23.34}        \\ \hline
		\textbf{ITR standard}                & 62.33       & 48.31       & 27.61       & 1.74        & 35.00        & 54.52       & 44.08       & 16.12       & 9.91        & 31.16        & 40.93       & 22.9        & 11.18       & 2.84        & 19.46        \\ \hline
		\rowcolor[HTML]{EFEFEF} 
		\textbf{Accuracy}                & 0.94        & 0.89        & 0.82        & 0.55        & 0.80         & 0.94        & 0.9         & 0.74        & 0.63        & 0.80         & 0.87        & 0.76        & 0.62        & 0.49        & 0.68         \\ \hline
		\textbf{Mean detection time}                & 1.14        & 1.22        & 1.56        & 1.72        & 1.41         & 1.33        & 1.37        & 1.9         & 1.61        & 1.55         & 1.33        & 1.41        & 1.29        & 1.56        & 1.4          \\ \hline
		\rowcolor[HTML]{EFEFEF} 
		\textbf{Number of predictions}            & 790         & 606         & 308         & 348         & 513          & 1008        & 857         & 270         & 433         & 642          & 995         & 718         & 1231        & 479         & 856          \\ \hline
	\end{tabular}
}
\end{table}

\begin{table}[]
	\def\arraystretch{1.5}
	\centering
	\caption{Comparison of proposed classification method, proposed method which always makes a prediction and random forest.}
	\label{tab:classification}
		\resizebox{\textwidth}{!}
	{
	\begin{tabular}{|l|l|l|l|l!{\vrule width 2pt}l!{\vrule width 2pt}l|l|l|l!{\vrule width 2pt}l!{\vrule width 2pt}l|l|l|l!{\vrule width 2pt}l!{\vrule width 2pt}}
		\hline
		& \multicolumn{5}{l|}{\textbf{Proposed method}}                        & \multicolumn{5}{l|}{\textbf{Proposed method: Always predicts}}       & \multicolumn{5}{l|}{\textbf{Random Forest}}                          \\ \cline{2-16} 
		\multirow{-2}{*}{\textbf{}} & \textbf{S1} & \textbf{S3} & \textbf{S3} & \textbf{S4} & \textbf{Avg} & \textbf{S1} & \textbf{S2} & \textbf{S3} & \textbf{S4} & \textbf{Avg} & \textbf{S1} & \textbf{S2} & \textbf{S3} & \textbf{S4} & \textbf{Avg} \\ \hline
		\rowcolor[HTML]{EFEFEF} 
		\textbf{ITR proposed}               & 64.62       & 53.73       & 33.19       & 5.65        & \textbf{39.30}        & 41.91       & 31.62       & 19.32       & 1.8         & \textbf{23.67}        & 40.4        & 31.01       & 19.08       & 2.46        & \textbf{23.24}        \\ \hline
		\textbf{ITR standard}                & 62.33       & 48.31       & 27.61       & 1.74        & 35.00        & 38.31       & 26.94       & 13.12       & 0.02        & 19.60        & 38.63       & 26.03       & 15.62       & 0.97        & 20.31        \\ \hline
		\rowcolor[HTML]{EFEFEF} 
		\textbf{Accuracy}                & 0.94        & 0.89        & 0.82        & 0.55        & 0.8          & 0.79        & 0.72        & 0.6         & 0.35        & 0.62         & 0.79        & 0.71        & 0.63        & 0.48        & 0.65         \\ \hline
		\textbf{Mean detection time}                & 1.14        & 1.22        & 1.56        & 1.72        & 1.41         & 1           & 1           & 1           & 1           & 1            & 1           & 1           & 1           & 1           & 1            \\ \hline
		\rowcolor[HTML]{EFEFEF} 
		\textbf{Number of predictions}            & 790         & 606         & 308         & 348         & 513          & 1695        & 1695        & 1695        & 1695        & 1695         & 1695        & 1695        & 1695        & 1695        & 1695         \\ \hline
	\end{tabular}
}
\end{table}

\begin{table}[h!]
	\def\arraystretch{1.5}
	\centering
	\caption{Proposed classifier compared to Random Forest and related work.}
	\label{table}
	\resizebox{\textwidth}{!}
	{
	\begin{tabular}{|l|l|l|l|l|l|l|l|l|}
		\hline
		& \textbf{\begin{tabular}[c]{@{}l@{}}Feature\\ extraction\end{tabular}}                    & \textbf{Classification}                                                           & \textbf{Classes}          & \textbf{\begin{tabular}[c]{@{}l@{}}Window\\ length (s)\end{tabular}} & \textbf{MDT (s)}             & \textbf{Accuracy}            & \textbf{\begin{tabular}[c]{@{}l@{}}ITR\\ standard\end{tabular}} & \textbf{\begin{tabular}[c]{@{}l@{}}ITR\\ proposed\end{tabular}} \\ \hline
		& \cellcolor[HTML]{EFEFEF}PSDA + CCA & \cellcolor[HTML]{EFEFEF}Proposed method & \cellcolor[HTML]{EFEFEF}3 & \cellcolor[HTML]{EFEFEF}1                                            & \cellcolor[HTML]{EFEFEF}1.41 & \cellcolor[HTML]{EFEFEF}80\% & \cellcolor[HTML]{EFEFEF}\textbf{35.00}                                   & \cellcolor[HTML]{EFEFEF}\textbf{39.30}                                   \\ \cline{2-9} 
		& PSDA + CCA                                                                               & Random Forest                                                                     & 3                         & 1                                                                    & 1                            & 64\%                         & 20.46                                                           & 23.73                                                           \\ \cline{2-9} 
		& \cellcolor[HTML]{EFEFEF}PSDA                                                             & \cellcolor[HTML]{EFEFEF}Proposed  method & \cellcolor[HTML]{EFEFEF}3 & \cellcolor[HTML]{EFEFEF}1                                            & \cellcolor[HTML]{EFEFEF}1.4  & \cellcolor[HTML]{EFEFEF}68\% & \cellcolor[HTML]{EFEFEF}\textbf{19.46}                                   & \cellcolor[HTML]{EFEFEF}\textbf{23.34}                                   \\ \cline{2-9} 
		\multirow{-4}{*}{\textbf{\begin{tabular}[c]{@{}l@{}}This\\ work\end{tabular}}} & PSDA                                                                                     & Random Forest                                                                     & 3                         & 1                                                                    & 1                            & 53\%                         & 9.64                                                            & 12.41                                                           \\ \hline
		\textbf{\cite{7_bifb_psda_cca}}                                                               & \cellcolor[HTML]{EFEFEF}BIFB                                                             & \cellcolor[HTML]{EFEFEF}No ML                                                     & \cellcolor[HTML]{EFEFEF}3 & \cellcolor[HTML]{EFEFEF}N/A                                          & \cellcolor[HTML]{EFEFEF}8.4  & \cellcolor[HTML]{EFEFEF}88\% & \cellcolor[HTML]{EFEFEF}8.2                                     & \cellcolor[HTML]{EFEFEF}-                                       \\ \hline
		\textbf{\cite{11_fft_bilinear}}                                                               & PSDA                                                                                     & Bilinear separation                     & 2                         & 1                                                                    & N/A                          & 74\%                         & 16                                                              & -                                                               \\ \hline
	\end{tabular}}
\end{table}

\section{Discussion}

From the results it can be seen that allowing the classifier to not make predictions when it is not confident enough results in better performance. This can be seen as a trade-off between accuracy and mean detection time: allowing the classifier to not make predictions results in better accuracy but worse mean detection time. According to ITR, which is a combination of both of these measures, allowing to not make predictions results in better performance. Fair comparison between different methods was possible thanks to the proposed classification method, which tries to find optimal discrimination threshold for each of the methods and thus the results are not affected by manual parameter selection. 

The proposed method outperforms both, the Random Forest classifier and the previously published comparable results in terms of ITR. If one values accuracy over ITR, the performance measure can be modified to take this into account and therefore a classifier with higher accuracy can be obtained. By combining multiple feature extraction methods, the achieved ITR is more than two times larger than the ones reported in related work. For one of the subjects, ITR of over 60 bit/min was achieved in this work. For further comparison to related work see~\cite{magister}.

For future work, the classifier will be tested online as the functionality for doing that is already implemented in our software. This is required because in the current work online ITR and MDT were estimated and might differ from the actual values. Online testing should also provide information whether the learned discrimination thresholds can be used in long periods of time or the thresholds need to be re-learned frequently, which would decrease usability of this method. In our offline experiments, however, the proposed method clearly outperforms the previous comparable results.

\section{Conclusion}

The main conclusions of the work are the following
\begin{itemize}
	\item Proposed classifier outperforms previous results (Table~\ref{table}).
	\item Combining feature extraction methods outperforms the same methods used individually (Table~\ref{tab:extraction}).
	\item Allowing the classifier to not make predictions when it is not confident enough results in better performance (Tables~\ref{tab:classification}).
\end{itemize}
In this work, a classification method for SSVEP-based BCI based on ITR optimisation was proposed. This enabled fair comparison of different feature extraction methods by finding the best discrimination threshold for each method. This clearly gives better results than manually chosen thresholds, as has been done in previous works, and is faster than grid-search method. By combining the features of different feature extraction methods using LDA, the proposed classification method can be used with multiple feature extraction methods, which increases the overall performance. Finally, by allowing the classifier to not make predictions when it is not confident enough is shown to be a good way to decrease the proportion of false-positive predictions. 

\section{Acknowledgments}

The authors thank the financial support from The Estonian Research Council through the personal research grant PUT1476. We also acknowledge funding by the European Regional Development Fund through the Estonian Center of Excellence in IT, EXCITE.

\section{References}
\bibliographystyle{bib-style}
\bibliography{references}

\providecommand{\newblock}{}
\begin{thebibliography}{10}
\expandafter\ifx\csname url\endcsname\relax
  \def\url#1{{\tt #1}}\fi
\expandafter\ifx\csname urlprefix\endcsname\relax\def\urlprefix{URL }\fi
\providecommand{\eprint}[2][]{\url{#2}}

\bibitem{ssvep_stim}
Zhu D, Bieger J, Molina G~G and Aarts R~M 2010 A survey of stimulation methods
  used in {SSVEP}-based {BCI}s {\em Computational Intelligence and
  Neuroscience\/} {\bf 2010} 1--13

\bibitem{itr_wolpaw}
Wolpaw J~R, Ramoser H, McFarland D~J and Pfurtscheller G 1998 {EEG-Based
  Communication: Improved Accuracy by Response Verification} {\em IEEE
  Transactions on Rehabilitation Engineering\/} {\bf 6} 326--333

\bibitem{11_fft_bilinear}
Jukiewicz M and Cysewska-Sobusiak A 2015 {Implementation of Bilinear Separation
  algorithm as a classification method for SSVEP-based brain-computer
  interface} {\em Measurement Automation Monitoring\/} {\bf 61} 51--53

\bibitem{12_amuse_ica}
Karnati V~B~R, Verma U~S~G, Amerineni J~S and Shah V~H 2014 {Frequency
  detection in Medium and High frequency SSVEP based Brain Computer Interface
  systems by scaling of sine-curve fit amplitudes} {\em {2014 First
  International Conference on Networks Soft Computing (ICNSC2014)}\/} pp
  300--303

\bibitem{1_music}
Velchev Y, Radev D and Radeva S 2016 {Features Extraction Based on Subspace
  Methods with Application to SSVEP BCI} {\em International Journal of Emerging
  Engineering Research and Technology\/} {\bf 4} 52--58

\bibitem{4_fft_svm}
Anindya S~F, Rachmat H~H and Sutjiredjeki E 2016 {A prototype of SSVEP-based
  BCI for home appliances control} {\em {2016 1st International Conference on
  Biomedical Engineering (IBIOMED)}\/} pp 1--6

\bibitem{7_bifb_psda_cca}
Demir A~F, Arslan H and Uysal I 2016 {Bio-inspired Filter Banks for SSVEP-based
  Brain-computer Interfaces} {\em {2016 IEEE International Conference on
  Biomedical and Health Informatics (BHI)}\/} (Las Vegas, NV, USA)

\bibitem{dataset}
Bakardjian H, Tanaka T and Cichocki A 2010 {Optimization of SSVEP brain
  responses with application to eight-command Brain--Computer Interface} {\em
  Neuroscience letters\/} {\bf 469} 34--38

\bibitem{psda}
Cheng M, Gao X, Gao S and Xu D 2002 Design and implementation of a
  brain-computer interface with high transfer rates {\em Biomedical
  Engineering, IEEE Transactions on\/} {\bf 49} 1181--1186

\bibitem{cca}
Lin Z, Zhang C, Wu W and Gao X 2007 {Frequency Recognition Based on Canonical
  Correlation Analysis for SSVEP-based BCIs} {\em Biomedical Engineering\/}
  {\bf 54}

\bibitem{mec}
Friman O, Volosyak I and Graser A 2007 Multiple channel detection of
  steady-state visual evoked potentials for brain-computer interfaces {\em
  Biomedical Engineering, IEEE Transactions on\/} {\bf 54}(4) 742--750

\bibitem{LRT}
Zhang Y, Dong L, Zhang R, Yao D, Zhang Y and Xu P 2014 {An Efficient Frequency
  Recognition Method Based on Likelihood Ratio Test for SSVEP-Based BCI} {\em
  Computational and Mathematical Methods in Medicine\/} {\bf 2014}

\bibitem{cwt}
Zhang Z, Li X and Deng Z 2010 {A CWT-based SSVEP classification method for
  brain-computer interface system} {\em {2010 International Conference on
  Intelligent Control and Information Processing}\/} pp 43--48

\bibitem{magister}
Ingel A 2017 {\em {Machine Learning in VEP-based BCI}\/} Master's thesis
  University of Tartu

\bibitem{13_pca}
Yehia A~G, Eldawlatly S and Taher M 2015 {Principal component analysis-based
  spectral recognition for SSVEP-based Brain-Computer Interfaces} {\em {2015
  Tenth International Conference on Computer Engineering Systems (ICCES)}\/} pp
  410--415

\end{thebibliography}

\newpage
\appendix
\section{Related work}\label{related_work}


\begin{table}[h!]
	\centering
	\def\arraystretch{1.2}
	\caption{Results of the related articles~\cite{magister}. In each of these works the dataset by Bakardijan \textit{et al.}~\cite{dataset} was used.}
	\label{tab:related_work_table}
	\resizebox{0.9\textwidth}{!}{\begin{tabular}{|l|Hl|l|l|l|l|l|l|l|}
		\hline
		\textbf{Article}                                       & \textbf{Section}                                      & \begin{tabular}[c]{@{}l@{}}\textbf{Feature}\\ \textbf{extraction}\end{tabular} & \textbf{Classification}                                                                                                                               & \textbf{Preprocessing}                                                                                        & \textbf{Classes}                                                                          & \begin{tabular}[c]{@{}l@{}}\textbf{Window}\\ \textbf{length/}\\ \textbf{MDT (s)}\end{tabular}                                    & \textbf{Subjects}                                                                & \textbf{Channels}                                                                                                                                    & \begin{tabular}[c]{@{}l@{}}\textbf{Accuracy}\\ \textbf{(\%)}\end{tabular} \\ \hline
		\multirow{4}{*}{\cite{1_music}}            & \multirow{4}{*}{\ref{sec:1_music}}        & \multirow{4}{*}{MUSIC}                                       & \multirow{4}{*}{\begin{tabular}[c]{@{}l@{}}SVM,\\ Gaussian\\ kernel,\\ one-vs-one\end{tabular}}                                              & \multirow{4}{*}{N/A}                                                                                 & \multirow{4}{*}{\begin{tabular}[c]{@{}l@{}}8 Hz,\\ 14 Hz\end{tabular}}           & 0.5                                                                                                 & \multirow{4}{*}{\begin{tabular}[c]{@{}l@{}}All\end{tabular}}   & \multirow{4}{*}{N/A}                                                                                                                        & 63.2                                                    \\ \cline{7-7} \cline{10-10} 
		&                                              &                                                              &                                                                                                                                              &                                                                                                      &                                                                                  & 1                                                                                                   &                                                                         &                                                                                                                                             & 75                                                      \\ \cline{7-7} \cline{10-10} 
		&                                              &                                                              &                                                                                                                                              &                                                                                                      &                                                                                  & 2                                                                                                   &                                                                         &                                                                                                                                             & 91.1                                                    \\ \cline{7-7} \cline{10-10} 
		&                                              &                                                              &                                                                                                                                              &                                                                                                      &                                                                                  & 4                                                                                                   &                                                                         &                                                                                                                                             & 94.3                                                    \\ \hline
		\multirow{3}{*}{\cite{4_fft_svm}}         & \multirow{3}{*}{\ref{4_fft_svm}}         & \multirow{3}{*}{PSDA}                                        & SVM linear                                                                                                                                   & \multirow{3}{*}{\begin{tabular}[c]{@{}l@{}}Windowed\\ sinc filter,\\ Blackman\\ window\end{tabular}} & \multirow{3}{*}{\begin{tabular}[c]{@{}l@{}}8 Hz,\\ 14 Hz,\\ 28 Hz\end{tabular}}  & \multirow{3}{*}{N/A}                                                                                & \multirow{3}{*}{\begin{tabular}[c]{@{}l@{}}All\end{tabular}}   & \multirow{3}{*}{\begin{tabular}[c]{@{}l@{}}O1, O2,\\ POz, Oz\end{tabular}}                                                                  & 65                                                    \\&&&&&&&&&\\ \cline{4-4} \cline{10-10} 
		&                                              &                                                              & SVM RBF                                                                                                                                      &                                                                                                      &                                                                                  &                                                                                                     &                                                                         &                                                                                                                                             & 71.67               \\&&&&&&&&&\\ \hline
		\multirow{12}{*}{\cite{7_bifb_psda_cca}} & \multirow{12}{*}{\ref{7_bifb_psda_cca}} & \multirow{4}{*}{BIFB}                                        & \multirow{12}{*}{\begin{tabular}[c]{@{}l@{}}No machine\\ learning\\ (but requires\\ finding\\ subject\\ specific\\ parameters)\end{tabular}} & \multirow{8}{*}{\begin{tabular}[c]{@{}l@{}}Bandpass\\ filter,\\ Hamming\\ window\end{tabular}}       & \multirow{12}{*}{\begin{tabular}[c]{@{}l@{}}8 Hz,\\ 14 Hz,\\ 28 Hz\end{tabular}} & 7.5                                                                                                 & 1                                                                       & \multirow{12}{*}{Oz}                                                                                                                        & 100                                                     \\ \cline{7-8} \cline{10-10} 
		&                                              &                                                              &                                                                                                                                              &                                                                                                      &                                                                                  & 7.8                                                                                                 & 2                                                                       &                                                                                                                                             & 100                                                     \\ \cline{7-8} \cline{10-10} 
		&                                              &                                                              &                                                                                                                                              &                                                                                                      &                                                                                  & 10                                                                                                  & 3                                                                       &                                                                                                                                             & 86.7                                                    \\ \cline{7-8} \cline{10-10} 
		&                                              &                                                              &                                                                                                                                              &                                                                                                      &                                                                                  & 8.33                                                                                                & 4                                                                       &                                                                                                                                             & 66.7                                                    \\ \cline{3-3} \cline{7-8} \cline{10-10} 
		&                                              & \multirow{4}{*}{PSDA}                                        &                                                                                                                                              &                                                                                                      &                                                                                  & 10                                                                                                  & 1                                                                       &                                                                                                                                             & 66.7                                                    \\ \cline{7-8} \cline{10-10} 
		&                                              &                                                              &                                                                                                                                              &                                                                                                      &                                                                                  & 9                                                                                                   & 2                                                                       &                                                                                                                                             & 66.7                                                    \\ \cline{7-8} \cline{10-10} 
		&                                              &                                                              &                                                                                                                                              &                                                                                                      &                                                                                  & 15                                                                                                  & 3                                                                       &                                                                                                                                             & 60                                                      \\ \cline{7-8} \cline{10-10} 
		&                                              &                                                              &                                                                                                                                              &                                                                                                      &                                                                                  & 15                                                                                                  & 4                                                                       &                                                                                                                                             & 6.7                                                     \\ \cline{3-3} \cline{5-5} \cline{7-8} \cline{10-10} 
		&                                              & \multirow{4}{*}{CCA}                                         &                                                                                                                                              & \multirow{4}{*}{N/A}                                                                                 &                                                                                  & 4                                                                                                   & 1                                                                       &                                                                                                                                             & 73.3                                                    \\ \cline{7-8} \cline{10-10} 
		&                                              &                                                              &                                                                                                                                              &                                                                                                      &                                                                                  & 4                                                                                                   & 2                                                                       &                                                                                                                                             & 60                                                      \\ \cline{7-8} \cline{10-10} 
		&                                              &                                                              &                                                                                                                                              &                                                                                                      &                                                                                  & 5                                                                                                   & 3                                                                       &                                                                                                                                             & 66.7                                                    \\ \cline{7-8} \cline{10-10} 
		&                                              &                                                              &                                                                                                                                              &                                                                                                      &                                                                                  & 3                                                                                                   & 4                                                                       &                                                                                                                                             & 66.7                                                    \\ \hline
		\multirow{4}{*}{\cite{11_fft_bilinear}}   & \multirow{4}{*}{\ref{11_fft_bilinear}}   & \multirow{4}{*}{PSDA}                                        & \multirow{2}{*}{\begin{tabular}[c]{@{}l@{}}Bilinear\\ separation\end{tabular}}                                                               & \multirow{4}{*}{N/A}                                                                                 & \multirow{4}{*}{\begin{tabular}[c]{@{}l@{}}8 Hz,\\ 14 Hz\end{tabular}}           & 1                                                                                                   & 1                                                                       & \multirow{4}{*}{\begin{tabular}[c]{@{}l@{}}O1, O2,\\ Oz\end{tabular}}                                                                       & 93                                                      \\ \cline{7-8} \cline{10-10} 
		&                                              &                                                              &                                                                                                                                              &                                                                                                      &                                                                                  & 1                                                                                                   & \begin{tabular}[c]{@{}l@{}}All\end{tabular}                    &                                                                                                                                             & $\sim$74                                                \\ \cline{4-4} \cline{7-8} \cline{10-10} 
		&                                              &                                                              & \multirow{2}{*}{SVM}                                                                                                                         &                                                                                                      &                                                                                  & 1                                                                                                   & 1                                                                       &                                                                                                                                             & 90                                                      \\ \cline{7-8} \cline{10-10} 
		&                                              &                                                              &                                                                                                                                              &                                                                                                      &                                                                                  & 1                                                                                                   & \begin{tabular}[c]{@{}l@{}}All\end{tabular}                    &                                                                                                                                             & $\sim$71                                                \\ \hline
		\multirow{3}{*}{\cite{12_amuse_ica}}      & \multirow{3}{*}{\ref{12_amuse_ica}}      & \multirow{3}{*}{\begin{tabular}[c]{@{}l@{}}Least\\ square\\ sine  fitting\end{tabular}}                   & \multirow{3}{*}{\begin{tabular}[c]{@{}l@{}}No machine\\ learning\end{tabular}}                                                               & \multirow{3}{*}{AMUSE}                                                                               & 8 Hz                                                                             & \multirow{3}{*}{0.5}                                                                                & \multirow{3}{*}{\begin{tabular}[c]{@{}l@{}}Two\\ subjects\end{tabular}} & \multirow{3}{*}{\begin{tabular}[c]{@{}l@{}}8 subject\\ specific\\ channels\end{tabular}}                                                    & 86                                                      \\ \cline{6-6} \cline{10-10} 
		&                                              &                                                              &                                                                                                                                              &                                                                                                      & 14 Hz                                                                            &                                                                                                     &                                                                         &                                                                                                                                             & 83                                                      \\ \cline{6-6} \cline{10-10} 
		&                                              &                                                              &                                                                                                                                              &                                                                                                      & 28 Hz                                                                            &                                                                                                     &                                                                         &                                                                                                                                             & 92                                                      \\ \hline
		\multirow{16}{*}{\cite{13_pca}}            & \multirow{16}{*}{\ref{13_pca}}            & \multirow{4}{*}{\begin{tabular}[c]{@{}l@{}}PSDA\\+ PCA\end{tabular}}                                  & \multirow{16}{*}{\begin{tabular}[c]{@{}l@{}}LDA (and\\ grid search\\ for subject\\ specific\\ parameters)\end{tabular}}                      & \multirow{16}{*}{\begin{tabular}[c]{@{}l@{}}Bandbass\\ filter, CAR,\\ MAF\end{tabular}}              & \multirow{16}{*}{\begin{tabular}[c]{@{}l@{}}8 Hz,\\ 14 Hz,\\ 28 Hz\end{tabular}} & \multirow{16}{*}{\begin{tabular}[c]{@{}l@{}}Average\\ over time\\ windows\\ up to 4 s\end{tabular}} & 1                                                                       & \multirow{16}{*}{\begin{tabular}[c]{@{}l@{}}Subject\\ specific\\ from the\\ selection:\\ P7, P3,\\ Pz, P4,\\ P8, O1,\\ Oz, O2\end{tabular}} & 92.43                                                   \\ \cline{8-8} \cline{10-10} 
		&                                              &                                                              &                                                                                                                                              &                                                                                                      &                                                                                  &                                                                                                     & 2                                                                       &                                                                                                                                             & 87.39                                                   \\ \cline{8-8} \cline{10-10} 
		&                                              &                                                              &                                                                                                                                              &                                                                                                      &                                                                                  &                                                                                                     & 3                                                                       &                                                                                                                                             & 94.56                                                   \\ \cline{8-8} \cline{10-10} 
		&                                              &                                                              &                                                                                                                                              &                                                                                                      &                                                                                  &                                                                                                     & 4                                                                       &                                                                                                                                             & 76.61                                                   \\ \cline{3-3} \cline{8-8} \cline{10-10} 
		&                                              & \multirow{4}{*}{\begin{tabular}[c]{@{}l@{}}Multi-\\way\\CCA\end{tabular}}                               &                                                                                                                                              &                                                                                                      &                                                                                  &                                                                                                     & 1                                                                       &                                                                                                                                             & 80.50                                                   \\ \cline{8-8} \cline{10-10} 
		&                                              &                                                              &                                                                                                                                              &                                                                                                      &                                                                                  &                                                                                                     & 2                                                                       &                                                                                                                                             & 71.16                                                   \\ \cline{8-8} \cline{10-10} 
		&                                              &                                                              &                                                                                                                                              &                                                                                                      &                                                                                  &                                                                                                     & 3                                                                       &                                                                                                                                             & 76.76                                                   \\ \cline{8-8} \cline{10-10} 
		&                                              &                                                              &                                                                                                                                              &                                                                                                      &                                                                                  &                                                                                                     & 4                                                                       &                                                                                                                                             & 60.53                                                   \\ \cline{3-3} \cline{8-8} \cline{10-10} 
		&                                              & \multirow{4}{*}{\begin{tabular}[c]{@{}l@{}}Multi-\\set\\CCA\end{tabular}}                               &                                                                                                                                              &                                                                                                      &                                                                                  &                                                                                                     & 1                                                                       &                                                                                                                                             & 69.98                                                   \\ \cline{8-8} \cline{10-10} 
		&                                              &                                                              &                                                                                                                                              &                                                                                                      &                                                                                  &                                                                                                     & 2                                                                       &                                                                                                                                             & 58.35                                                   \\ \cline{8-8} \cline{10-10} 
		&                                              &                                                              &                                                                                                                                              &                                                                                                      &                                                                                  &                                                                                                     & 3                                                                       &                                                                                                                                             & 79.80                                                   \\ \cline{8-8} \cline{10-10} 
		&                                              &                                                              &                                                                                                                                              &                                                                                                      &                                                                                  &                                                                                                     & 4                                                                       &                                                                                                                                             & 47.17                                                   \\ \cline{3-3} \cline{8-8} \cline{10-10} 
		&                                              & \multirow{4}{*}{CCA}                                         &                                                                                                                                              &                                                                                                      &                                                                                  &                                                                                                     & 1                                                                       &                                                                                                                                             & 62.69                                                   \\ \cline{8-8} \cline{10-10} 
		&                                              &                                                              &                                                                                                                                              &                                                                                                      &                                                                                  &                                                                                                     & 2                                                                       &                                                                                                                                             & 59.52                                                   \\ \cline{8-8} \cline{10-10} 
		&                                              &                                                              &                                                                                                                                              &                                                                                                      &                                                                                  &                                                                                                     & 3                                                                       &                                                                                                                                             & 65.51                                                   \\ \cline{8-8} \cline{10-10} 
		&                                              &                                                              &                                                                                                                                              &                                                                                                      &                                                                                  &                                                                                                     & 4                                                                       &                                                                                                                                             & 56.04                                                   \\ \hline
	\end{tabular}}
\end{table}

\section{Skew normal distribution fit to data}\label{distribution}

\begin{figure}[h!]
	\centering
	\input{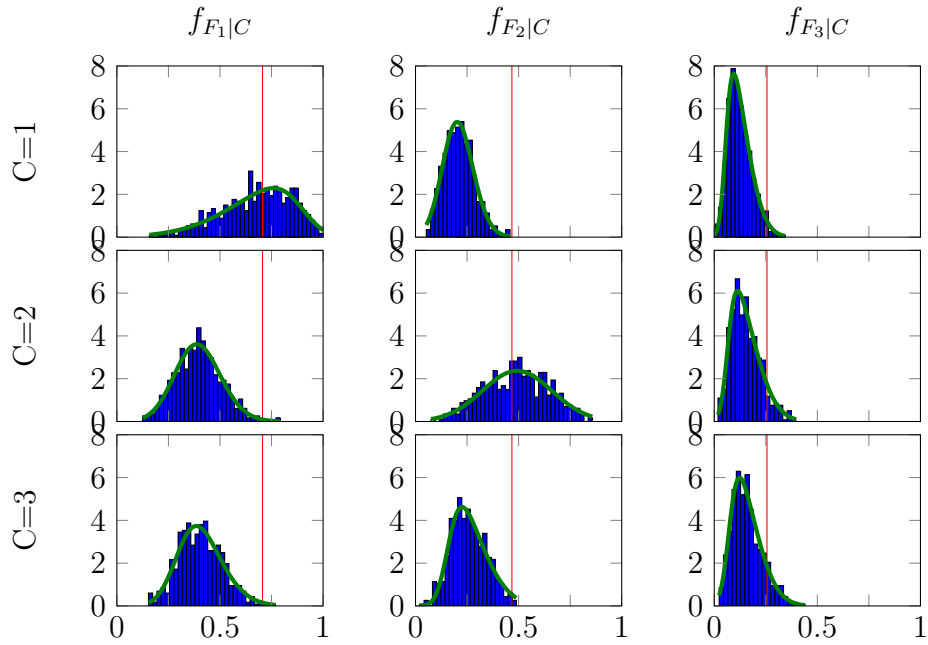}
	\caption{Histograms of features given class (blue) for subject 1. Green line denotes skew normal distribution fit to the data and red lines indicate possible thresholds~\cite{magister}.}
	\label{fig:class_figure5}
\end{figure}
~
\begin{figure}[h!]
	\centering
	\input{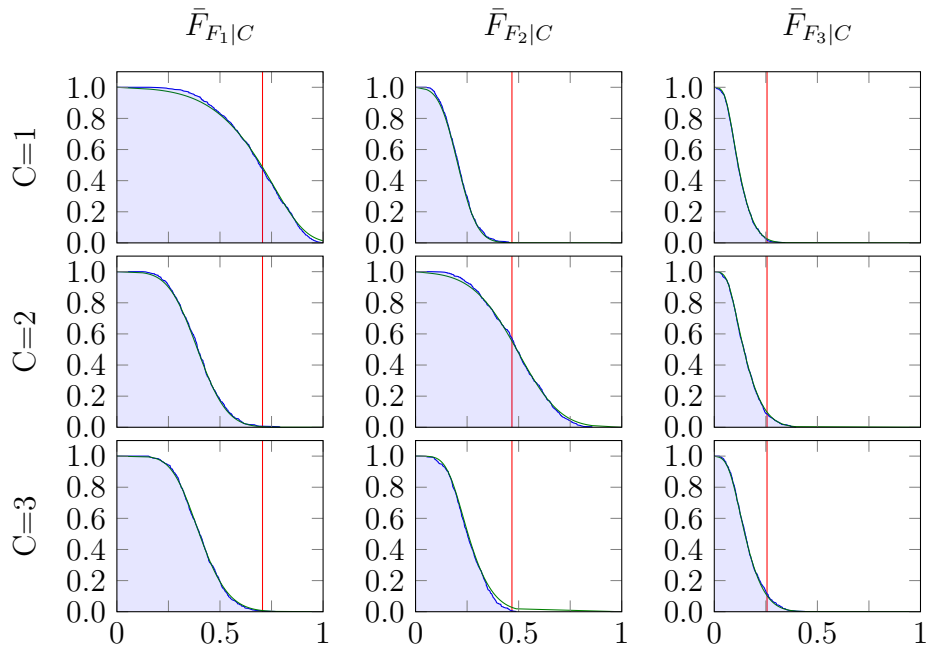}
	\caption{Complementary cumulative distribution function of the corresponding probability density function in Figure~\ref{fig:class_figure5}~\cite{magister}.}
	\label{fig:class_figure7}
\end{figure}

\section{ITR calculation when predictions can be skipped}\label{appendix_conditioning}

In this appendix, we derive the formula for calculating ITR taking into account the fact that the classifier might not make a prediction at each time step. Therefore, in addition to the events of prediction class $i$ and correct class being $j$ denoted by $P_i$ and $C_j$ respectively, let us denote the event of making a prediction as $M$. Furthermore, let us assume that the number of classes is $n$. Then
\begin{align}
	M=\bigcup_{i=1}^nP_i
\end{align}
and since predicting different classes are mutually exclusive events, it holds that
\begin{align}
	\mathbf{P}(M)=\mathbf{P}\left(\bigcup_{i=1}^nP_i\right)=\sum_{i=1}^{n}\mathbf{P}(P_i)
\end{align}
where $\mathbf{P}(P_i)$ can be calculated using equation \eqref{Pi}. The previous probability is required to calculate
\begin{align}
	\mathbf{P}(P_i\mid M)=\frac{\mathbf{P}(P_i\cap M)}{\mathbf{P}(M)}= \frac{\mathbf{P}(P_i)}{\mathbf{P}(M)}
\end{align}
which is needed to calculate ITR. We also need to condition $C_j$ on the event $M$
\begin{align}
\mathbf{P}(C_j\mid M)
&=\frac{\mathbf{P}\left(C_j\cap\left(\bigcup_{i=1}^n P_i\right)\right)}{\mathbf{P}(M)}\nonumber\\
&=\frac{\mathbf{P}\left(\bigcup_{i=1}^n\left(C_j\cap P_i\right)\right)}{\mathbf{P}(M)}\nonumber\\
&=\frac{1}{\mathbf{P}(M)}\sum_{i=1}^{n}\mathbf{P}(C_j\cap P_i)\nonumber\\
&=\frac{1}{\mathbf{P}(M)}\sum_{i=1}^{n}\mathbf{P}(P_i\mid C_j)\mathbf{P}(C_j)
\end{align}
where $\mathbf{P}(C_j)$ can be calculated using equation \eqref{Cj}. The last probability needed to calculate ITR is given by
\begin{align}
\mathbf{P}(P_i\mid C_j,M)=\frac{\mathbf{P}(P_i\cap C_j\cap M)}{\mathbf{P}(C_j\cap M)}=\frac{\mathbf{P}(P_i\cap M\mid C_j)\mathbf{P}(C_j)}{\mathbf{P}(C_j\cap M)}=\frac{\mathbf{P}(P_i\mid C_j)}{\mathbf{P}(M\mid C_j)}
\end{align}
where
\begin{align}
\mathbf{P}(M\mid C_j)=\sum_{i=1}^{n}\mathbf{P}(P_i\mid C_j).
\end{align}
Finally, ITR can be calculated as
\begin{align}
I(P| M, C| M)=\sum_{i=1}^{n}\sum_{j=1}^{n}\mathbf{P}(P_i\mid C_j,M)\mathbf{P}(C_j\mid M) \log_2\left(\frac{\mathbf{P}(P_i\mid C_j,M)}{\mathbf{P}(P_j\mid M)}\right).
\end{align}
\section{Deriving gradient of ITR}\label{appendix_gradient}
In this appendix the formula for calculating gradient of ITR is derived. Let again feature value for class $i$ be modelled by random variable $F_i$ and denote the event of $k$ being the correct class by $C_k$. First note that probability density function $f_{F_i\mid C_k}$ of the random variable $F_i\mid C_k$ is the derivative of the corresponding cumulative density function $F_{F_i\mid C_k}$
\begin{align}
\frac{dF_{F_i\mid C_k}}{dt_i}=f_{F_i\mid C_k}.
\end{align}
This allows us to calculate
\begin{align}
\frac{\partial}{\partial t_{\ell}}\mathbf{P}(P_i\mid C_k)&=\left\{
\begin{array}{@{}ll@{}}
-f_{F_i\mid C_k}(t_i)\cdot\prod_{j\in\{1,\dots,n\}\setminus\{i\}}F_{F_j\mid C_k}(t_j) &\mbox{ if } i = \ell\\
\overline{F}_{F_i\mid C_k}(t_i)\cdot f_{F_\ell\mid C_k}(t_\ell)\cdot\prod_{j\in\{1,\dots,n\}\setminus\{i,\ell\}}F_{F_j\mid C_k}(t_j) &\mbox{ if } i\neq\ell
\end{array}\right.
\end{align}
that is needed later.

To make differentiation of ITR easier, the following property will be used
\begin{align}
	I(P|M, C|M)=H(P|M)-H((P|M)\mid (C|M))
\end{align}
where
\begin{align}
H(P|M)&=-\sum_{i=1}^{n}\mathbf{P}(P_i|M)\log_2\mathbf{P}(P_i|M)\\
H((P|M)\mid (C|M))&=\sum_{j=1}^{n}\mathbf{P}(C_j\mid M)H(P\mid C_j,M).
\end{align}
The corresponding partial derivatives are
\begin{align}
	\frac{\partial}{\partial t_\ell}H(P\mid M)&=-\sum_{i=1}^{n}\frac{\ln(\mathbf{P}(P_i\mid M))+1}{\ln2}\frac{\partial}{\partial t_\ell}\mathbf{P}(P_i\mid M)\\
	\frac{\partial}{\partial t_\ell}H((P|M)\mid (C|M))&=\sum_{j=1}^{n}\left(H(P\mid C_j,M)\frac{\partial}{\partial t_\ell}\mathbf{P}(C_k\mid M)+\mathbf{P}(C_k\mid M)\frac{\partial}{\partial t_\ell}H(P\mid C_j,M)\right)
\end{align}
and the partial derivatives for the probabilities are
\begin{align}
\frac{\partial}{\partial t_\ell}\mathbf{P}(P_i)&=\sum_{j=1}^{n}\mathbf{P}(C_j)\frac{\partial}{\partial t_\ell}\mathbf{P}(P_i\mid C_j)\\
\frac{\partial}{\partial t_\ell}\mathbf{P}(M)&=\sum_{j=1}^n \frac{\partial}{\partial t_\ell}\mathbf{P}(P_j)\\
\frac{\partial}{\partial t_\ell}\mathbf{P}(P_i\mid M) &=\frac{\mathbf{P}(M)\frac{\partial}{\partial t_\ell}\mathbf{P}(P_i)-\mathbf{P}(P_i)\frac{\partial}{\partial t_\ell}\mathbf{P}(M)}{\left(\mathbf{P}(M)\right)^2}\\
\frac{\partial}{\partial t_\ell}\mathbf{P}(C_k\mid M)&=\frac{ \left(\mathbf{P}(M)\sum_{j=1}^{n}\frac{\partial}{\partial t_\ell}\mathbf{P}(P_j\cap C_k)\right)- \left(\sum_{j=1}^{n}\mathbf{P}(P_j\cap C_k)\frac{\partial}{\partial t_\ell}\mathbf{P}(M)\right)
}{(\mathbf{P}(M))^2}\\
\frac{\partial}{\partial t_\ell}\mathbf{P}(M\mid C_j)&=\sum_{i=1}^{n}\frac{\partial}{\partial t_\ell}\mathbf{P}(P_i\mid C_j)
\end{align}
\begin{align}
\frac{\partial}{\partial t_\ell}\mathbf{P}(P_i\mid C_j,M)&= \frac{\mathbf{P}(M\mid C_j)\frac{\partial}{\partial t_\ell}\mathbf{P}(P_i\mid C_j)-\mathbf{P}(P_i\mid C_j)\frac{\partial}{\partial t_\ell}\mathbf{P}(M\mid C_j)}{\left(\mathbf{P}(M\mid C_j)\right)^2}.
\end{align}
Therefore we are able to calculate
\begin{align}
	\frac{\partial}{\partial t_\ell}I(P|M, C|M)&=\frac{\partial}{\partial t_\ell}H(P\mid M)-\frac{\partial}{\partial t_\ell}H(P\mid C,M).
\end{align}

Next, the partial derivative of MDT will be calculated
\begin{align}
	\frac{\partial}{\partial t_\ell}{\mbox{MDT}}&=-\frac{s}{(\mathbf{P}(M))^2}\cdot\frac{\partial}{\partial t_\ell}\mathbf{P}(M)
\end{align}
which together with the previous result can be used to get
\begin{align}
	\frac{\partial}{\partial t_\ell}\mbox{ITR}&=\left({\mbox{MDT}}\frac{\partial}{\partial t_\ell}I(P| M, C| M)-I(P| M, C| M)\frac{\partial}{\partial t_\ell}{\mbox{MDT}}\right)\frac{60}{{\mbox{MDT}}^2}.
\end{align}
Finally, the required gradient is
\begin{align}
	\nabla{\mbox{ITR}}=\left(\frac{\partial}{\partial t_1}{\mbox{ITR}}, \dots, \frac{\partial}{\partial t_n}{\mbox{ITR}}\right)^T.
\end{align}

\section{Implementation of the algorithm}\label{appendix_repo}
The implementation of the proposed classification algorithm can be found in Github repository\footnote{\url{https://github.com/antiingel/ITR-optimisation}}. The detailed instruction on how to run the scrip is in \texttt{README.md} file in the repository.

\end{document}